# On the Role of Separatrix Instabilities in Heating the Reconnection Outflow Region

M. Hesse[1,2], C. Norgren[1], P. Tenfjord[1], J. Burch[2], Y.-H. Liu[3], L.-J. Chen[4], N. Bessho[4], S. Wang[4], R. Nakamura[5], J. Eastwood[6], M. Hoshino[7], R. Torbert[8], R. Ergun[9]

[1]Birkeland Centre for Space Science, Department of Physics and Technology, University of Bergen, Norway. [2]Southwest Research Institute, San Antonio, TX, USA. [3]Dartmouth College, Hanover, NH, USA. [4]NASA, Goddard Space Flight Center, Greenbelt, MD, USA. [5]Space Research Institute, Austrian Academy of Sciences, Graz, Austria. [6]Imperial College, London, UK, [7]University of Tokyo, Tokyo, Japan, [8]University of New Hampshire, Durham, NH, USA, [9]University of Colorado, Boulder, CO, USA.

**ABSTRACT**

A study of the role of microinstabilities at the reconnection separatrix can play in heating the electrons during the transition from inflow to outflow is being presented. We find that very strong flow shears at the separatrix layer lead to counterstreaming electron distributions in the region around the separatrix, which become unstable to a beam-type instability. Similar to what has been seen in earlier research, the ensuing instability leads to the formation of propagating electrostatic solitons. We show here that this region of strong electrostatic turbulence is the predominant electron heating site when transiting from inflow to outflow. The heating is the result of heating generated by electrostatic turbulence driven by overlapping beams, and its macroscopic effect is a quasi-viscous contribution to the overall electron energy balance. We suggest that instabilities at the separatrix can play a key role in the overall electron energy balance in magnetic reconnection.

**I. INTRODUCTION**

Magnetic reconnection is arguably the most important transport and energy conversion process in collisionless plasmas[1,2]. It enables transport over large distances by means of a highly localized diffusion region, where, within different layers, ions and electrons become decoupled from the magnetic field. Particularly the physics of the smallest sub-region, the electron diffusion region, has been enigmatic for many years. Recent observations[3,4], however, have demonstrated that the laminar, thermal electron inertia-based (or, quasi-viscous) model[5,6] of its structure appears to be correct. The laminar nature of the diffusion region has been suggested to be the consequence of finite electron residence time[7].

The electron diffusion region is of crucial importance to the reconnection process, but, owing to its diminutive size, it cannot be the main actor in the overall energy conversion process. Energy conversion in magnetic reconnection occurs over macroscopic spatial scales, in the case of the Earth's magnetosphere over tens of Earth radii, whereas the typical dimensions of the electron diffusion region are a mere ten to 100km. Consequently, other processes have to come into play to facilitate large-scale energy conversion.

In magnetohydrodynamic (MHD) models this energy conversion is facilitated by slow shocks[8], or, a more general set of discontinuities in non-coplanar geometries[9] or in anisotropic plasmas[10]. In MHD, these discontinuities facilitate the conversion of incoming Poynting flux to enthalpy and kinetic energy flux[11]. Among other effects, the energy conversion process also provides for the pressure balance in the current layer by increasing plasma temperature and pressure to the level required to balance the magnetic pressure in the inflow region[12].

In a kinetic plasma, however, it is not a prior clear how this heating process works. For magnetized ions further down the outflow direction, the Cowley-Owen effect[13] of sweeping up populations at rest generates sufficient thermal energy if the counter-streaming ions beams are thermalized. For unmagnetized ions closer to the ion diffusion region, pick-up effects are likely to play a role in providing the required energization[14].

Outside of the electron diffusion region, electron heating is less understood. It is quite clear that the Cowley-Owen mechanism does not provide sufficient energy as the electron thermal speed is much larger than the Alfvén velocity. Similarly, an electron pickup effect will lead to particle velocities of, at most, an ion Alfvén speed, and hence not heat electron populations sufficiently to explain the observed ion/electron temperature ratios and to provide the electron contribution to the overall pressure balance. Consequently, other processes have to be at work to facilitate electron heating.

In this paper, we describe one such candidate process, and analyze, within a numerical model, its role in increasing the electron pressure. We show that this effect occurs naturally in the region surrounding the separatrix, and that it facilitates the major part of electron heating in the model. We suggest that processes such as the one discussed here can play a key role in the electron energy balance of a reconnecting current layer.

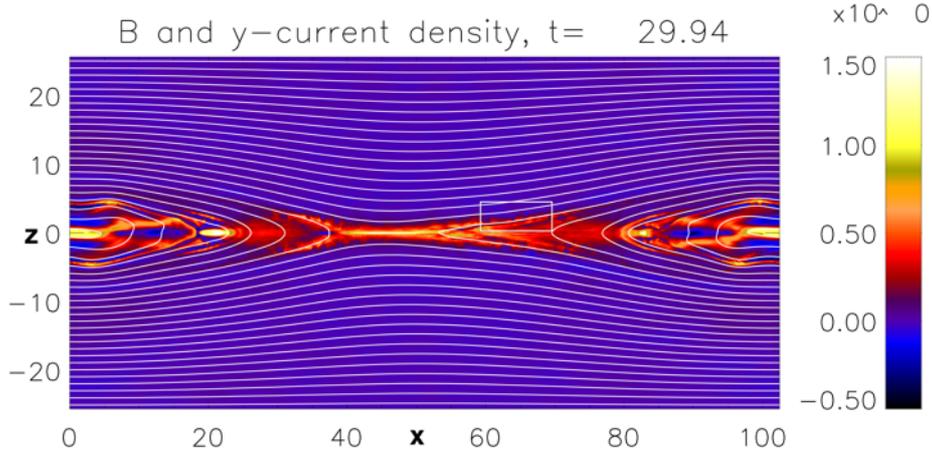

Figure 1. Magnetic field and out-of-plane current density at the time of the investigation. The white rectangle marks a sub-region for detailed analysis. There are noticeable fluctuations of the current density in the region along the separatrices.

## II. MODEL

The employed code is a 2.5D variant of our proven 3D particle-in-cell code[5], where periodic boundary conditions have been replaced by open boundaries, where normal derivatives of density, velocity, and isotropic pressure are assumed to vanish. Here and during the following chapter we normalize densities by a typical density $n_0$ in the current sheet, the magnetic field by the asymptotic value $B_0$ of the in-plane magnetic field. Ions are assumed to be protons (mass $m_p$) throughout, and length scales are normalized by the ion inertial length $c/\omega_i$, where the ion plasma frequency $\omega_i = \sqrt{e^2 n_0 / \varepsilon_0 m_p}$ is evaluated for the reference density. Velocities are measured in units of the ion Alfvén velocity $v_A = B_0 / \sqrt{\mu_0 m_p n_0}$ based on the reference magnitudes of magnetic field and density. The electric field is measured in units of $E_0 = v_A B_0$, and the pressure in units of $p_0 = B_0^2/\mu_0$. The above lengths scale and Ampere's law imply that the current density is normalized to $j_0 = \omega_i B_0 / c\mu_0$.

The initial condition is the same as in a recent investigation[7]. It consists of a poloidal

magnetic field, a modified Harris sheet[15] with a current sheet half width of half the ion

inertial length. The initial magnetic field is co-planar, i.e., does not contain a guide

field component. In a coordinate system where the *x* axis is in the direction of the

initial magnetic field, the *y* axis in the initial current direction, and the *z* direction

completes a right-handed coordinate system, the initial magnetic field is of the

following form:

$$B_x = \tanh(2z) + a_0 \pi/L_z \cos(2\pi x/L_x)\sin(\pi z/L_z)$$
(1a)

$$B_z = -a_0 2\pi/L_z \sin(2\pi x/L_x)\cos(\pi z/L_z)$$
(1b)

The perturbation amplitude $a_0$ is chosen to yield an initial value of the normal

magnetic field of about 3% of $B_0$. The system size is $L_x$=102.4 and $L_z$=51.2. The ion-

electron mass ratio is chosen to be 100. A total of $7 \times 10^{10}$ macro-particles are moved

on a 3200x3200 grid, with an electron/ion temperature ratio of $T_e/T_i$=0.2. The

simulation employs a ratio of electron plasma to electron cyclotron frequency of

$\omega_{pe}/\Omega_e$=2, and the ratio between the speed of light and the Alfven speed is therefore

$c/v_A$=20.

## III. SHEAR FLOW-DRIVEN INSTABILITY

Figure 1 shows the in-plane magnetic field and out-of-plane current density at the time of our investigation. A closer inspection reveals that there is considerable structure in the current density approximately at the location of the separatrix, particularly in an intermediate distance from the X-line. This structure is a clear indication that some kind of time-dependent, turbulent, process may be at work in this region. Figure 1 also shows a rectangle, which indicates a sub-region for more detailed analysis below.

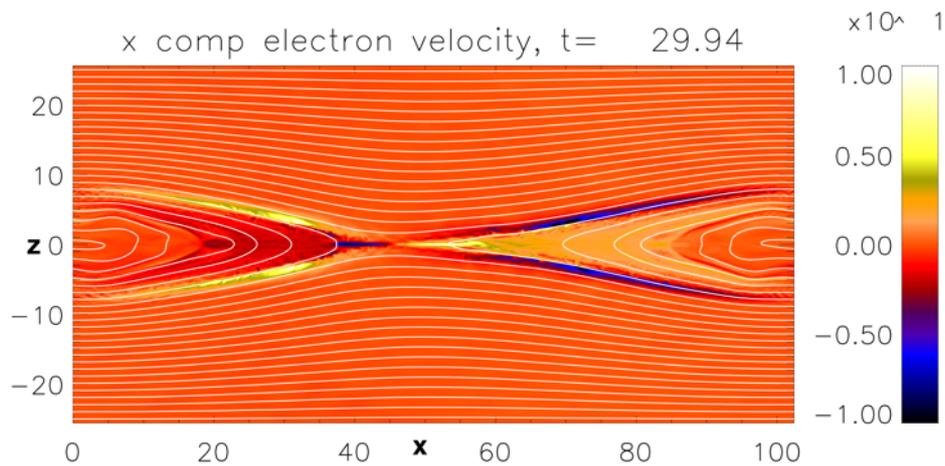

Figure 2. Magnetic field and x-component of the electron velocity. We find a fast jet directed toward the X-line just outside, or directly at, the separatrix location, and the broader reconnection outflow inside.

A suggestion for the source of this turbulence is shown in Figure 2, which displays the x-component of the electron flow velocity. As is apparent from the figure, there is a sharp gradient between the electron outflow inside the separatrix and the inflow, which is typically found to exist closer to the separatrix, or just outside of it[16]. It is interesting to note that the gradient scale length between the two adjacent flow jets is comparable to the local electron Larmor radius for a wide range of x values. Larmor-scale gradients imply that the two jets are not on well-separated magnetic flux tubes, but rather that a region of overlap exists because of finite Larmor radius effects. Such overlap regions can feature counter-streaming electron beams, which can be unstable to beam-type modes if the relative drift speed is large enough. Fig. 2 shows that the absolute difference between the bulk speeds of the two jets is already of the order to the electron Alfvén speed – hence instability conditions are likely to be found here. Indeed, a number of previous investigations have found instabilities and bipolar electric field in this region and suggested that they are closely tied to the reconnection process[17-20], and to electron heating[21].

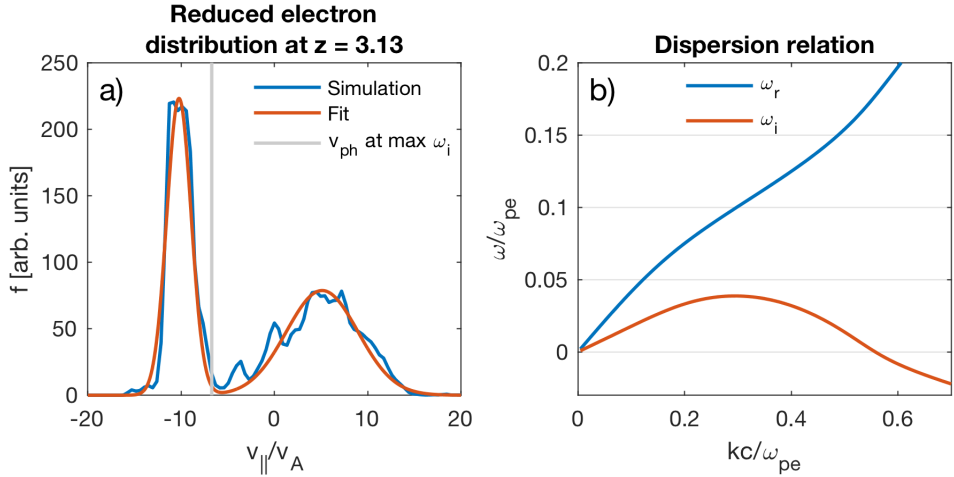

Figure 3. (a) One-dimensional reduced electron distribution as a function of the velocity parallel to the magnetic field at x=72.5 and z=3.13. The distribution shows the two-beam structure, which results from the mixing of the inflow and outflow populations. The blue line in the left panel is the distribution from the model, and the red line is the fit used to generate the dispersion solution on the right panel (b).

In order to demonstrate the existence of instabilities in our model, we calculate one-dimensional distributions as a function of the velocity parallel to the magnetic field at x=72.5. Figure 3a shows an example in the separatrix region at z = 3.13, where $v_\parallel<0$ (>0) corresponds to motion toward (away from) the X line. The distribution is found within a local density cavity and shows the two-beam structure, which results from the mixing of the inflow and outflow populations. Since the electric field structures are predominantly field aligned and propagate parallel to the magnetic field, we solve the one-dimensional electrostatic plasma dispersion equation to find the unstable wave modes associated with the distribution:

$$0 = 1 - \sum_s \frac{\omega_{ps}^2}{k^2 v_{ts}^2} Z'\left(\frac{\omega - k v_{ds}}{k v_{ts}}\right)$$

Here Z is the plasma dispersion function, $v_{ts} = (2k_B T_s/m_s)^{1/2}$ is the thermal speed, $\omega_{ps} = (n_s e^2/\epsilon_0 m_s)^{1/2}$ is the plasma frequency, and $v_{ds}$ is the drift speed of population *s*. We create a fit to the simulation distribution using two electron populations, with normalized parameters $v_{ts}$ = [1.8,5.4]$v_{A0}$, $v_{ds}$ = [-10.2,5.1]$v_A$, $\omega_{ps}$ = [0.060,0.179] $\omega_{pe}$ (Figure 3a). We note that the normalization is based on $B_0$ and $n_0$, and not the local values of B and n. The highest growth rate occurs for $kc/\omega_{pe}$ ~ 0.3 (Figure 3, right), corresponding to a wavelength lambda ~ 20 $c/\omega_{pe}$. In comparison, the wavelength in the model is about 1 $c/\omega_{pi}$ = 10$c/\omega_{pe}$. The phase velocity at maximum growth rate is $v_{ph}$ = -6.7$v_A$ as shown as a gray vertical line in Figure 3a. The propagation is therefore towards the X line, matching findings below. The results shown in Figure 3 are representative for a range of distributions found around the separatrix region.

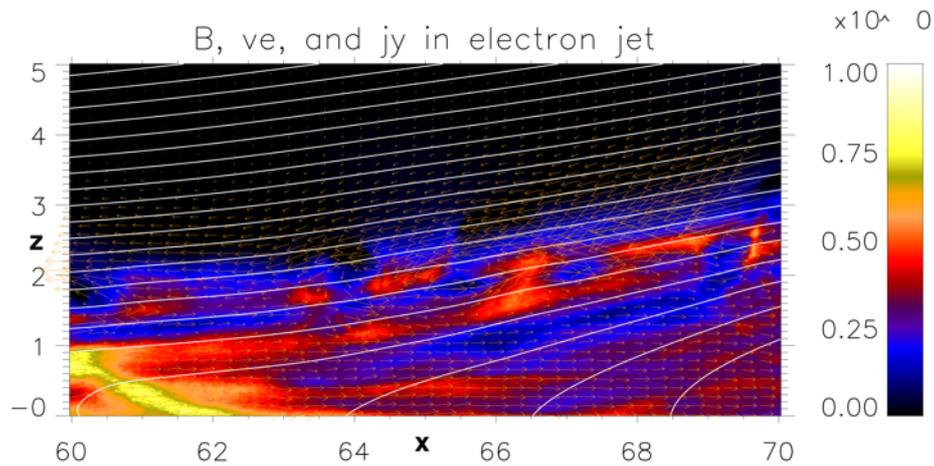

Figure 4. Blowup of a region around the upper right separatrix. Shown is magnetic field, out-of-plane current density, and in-plane electron flow vectors. There are vortical structures of the electron flow speed closely correlated with current density fluctuations.

The local magnetic field, current density, and in-plane electron flows are depicted in Figure 4. Here we find a substantially fractured current layer near the separatrix location, with apparent vortical deflections of the poloidal (in-plane) electron flow correlated with the current density structures. Vortical flow deflections of this type are likely to exist in conjunction with double layers[22-23], and they are an indication that major dissipation of the two flow layers may be in process.

Very strong bi-polar electric field structures exist indeed in the model, similar to previous models[17]. Figure 5 shows the x-component of the electric field in the region of investigation. It is apparent that these very strong electric fields – with amplitudes of a factor of more than ten times the reconnection electric field, exist in the spatial domain bridging the in- and outflow beams. Much of this electric field is parallel to the magnetic field, but there are substantial perpendicular fringe fields as well. The depths of the potential wells associated with the electric double structures are of the order of $\delta\phi \approx 0.4$, which implies that a significant amount of electrons is trapped.

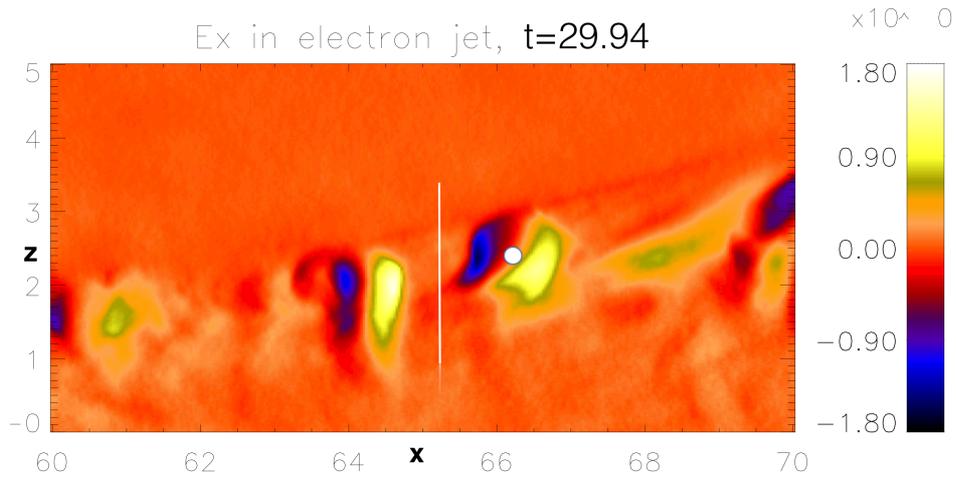

Figure 5. x-component of the electric field in the zoomed-in region. The figure demonstrates the existence of strong electric field double structures, with amplitudes an order of magnitude larger than the reconnection electric field. The vertical line and white dot indicate the locations distribution functions are calculated at (see below).

We can use two different simulation times to investigate the motion of these structures. This is shown in Figure 6. Here we see that the bipolar structure propagates toward the reconnection X-point, with a velocity of approximately half of the electron Alfvén speed. Similar motion directed toward the X-point has been seen before[19], but our model does not feature the whistler signatures found in that model, a likely consequence of the relatively slower propagation speed in the present simulation.

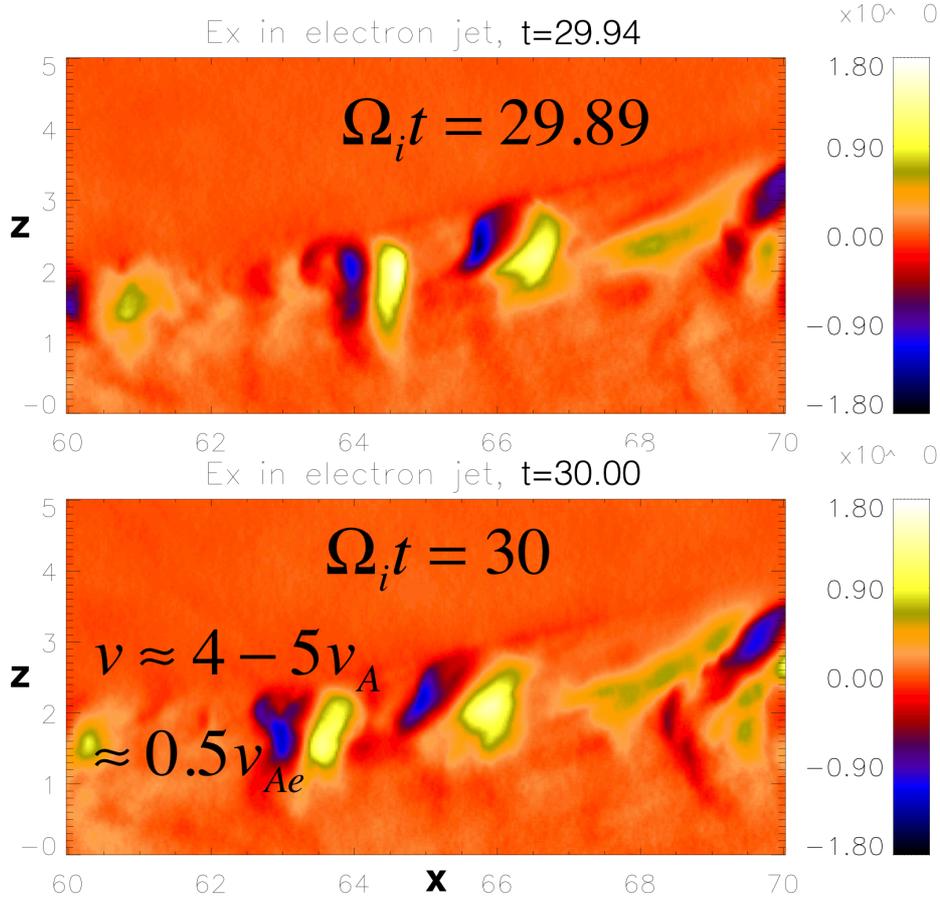

Figure 6. Time change of the position of the double layers. Evaluation of the motion toward the X point leads to propagation velocity of approximate half an electron Alfvén speed.

We now turn to an investigation of the distribution function structure in-between solitons. Figure 7 displays four reduced distributions $F(v_x, v_z)$ at x=65.25 and various levels of z. The location where these distributions are calculated is marked by the vertical line in fig. 5. Each distribution is based on collecting particles within a rectangle of dimensions $\Delta x \times \Delta z = 0.5 \times 0.1$. The relatively simple distribution in the inflow region at z=3.5 is replaced by a multi-beam distribution at z=2.7 – at a position approximately between the two electrostatic double structures in fig. 5. This latter distribution has an apparent average negative x-component of the velocity. Further inside the outflow region, the average motion becomes increasingly directed away from the X-line, with multiple striations in velocity space found like in previous investigations[24-26] closest to z=0. Overall, the distribution at z=1.1 has the highest temperature.

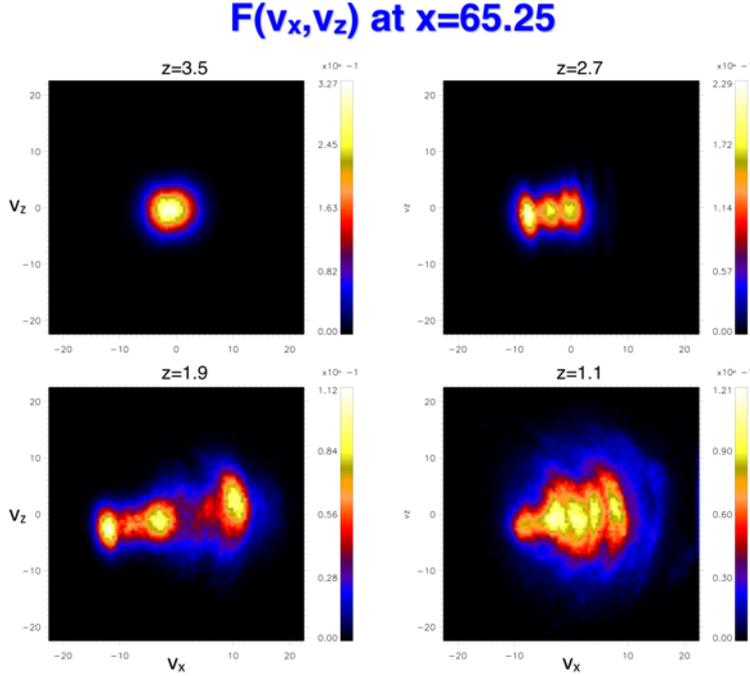

Figure 7: Reduced distributions F($v_x$,$v_z$) between two bipolar electric field structures. The distributions are taken at the indicated z locations. Furthest out, we find a simple inflow distribution, which is replaced by multi-beam distributions, and finally a partially thermalized distribution furthest inside. The likely source of the multibeam distributions structure is the Larmor-scale overlap between multiple beams in the outflow[24,26] and the inflowing distribution.

As expected from the arguments above, there is a trapped distribution inside the electrostatic structure. Figure 8, which features a distribution taken at the location marked by the white circle in fig. 5, shows the signatures of trapping: in the frame moving with the electric field structures, with the velocity marked by the arrow, we find counter-streaming particles bouncing between the potential walls. The majority of electrons appear to be trapped and at a specific energy level in the frame moving with the structure, but electrons at lower energy levels are trapped as well with lower phase space density levels.

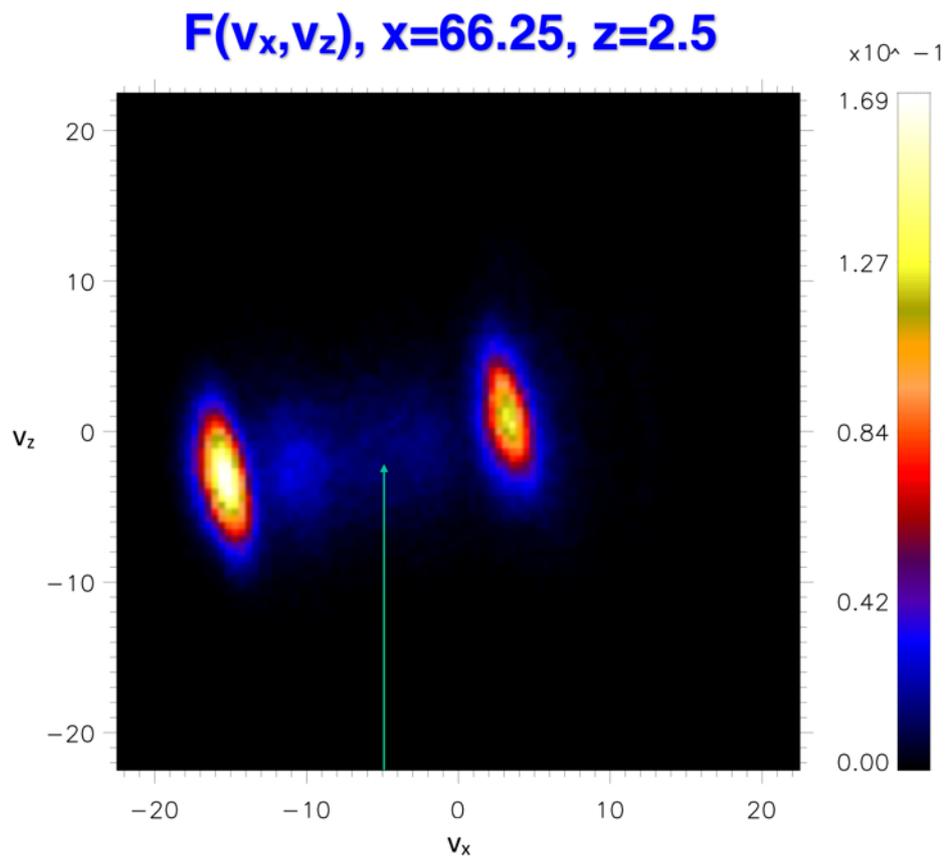

Figure 8: Reduced distributions $F(v_x,v_z)$ inside of the bipolar electric field structures. We find a typical trapped distribution, which features an average velocity of approximately an electron Alfvén speed.

Summarizing the results of this section, we find evidence of the nonlinear state of an electron beam-beam instability. The unstable situation appears to be created through finite Larmor radius effects, which overlap electrons streaming toward and away from the inner diffusion region. The instability leads to the formation of electrostatic double structures, which move against the outflow direction. In the following, we will now quantitatively investigate the effects this turbulence has on the overall electron energy budget.

## IV. ELECTRON HEATING

One of the key questions in magnetic reconnection research is related to how the magnetic energy in the inflow regions is converted to particle energy. There are a number of different aspects to this question, and one of them is focusing on the electrons, and, specifically, how a relatively tenuous inflow population can be energized sufficiently to turn into the outflow population. We will argue here that the interaction between strong, counter-streaming electron beams close to the separatrix region can, if leading to plasma instabilities, provide a significant contribution.

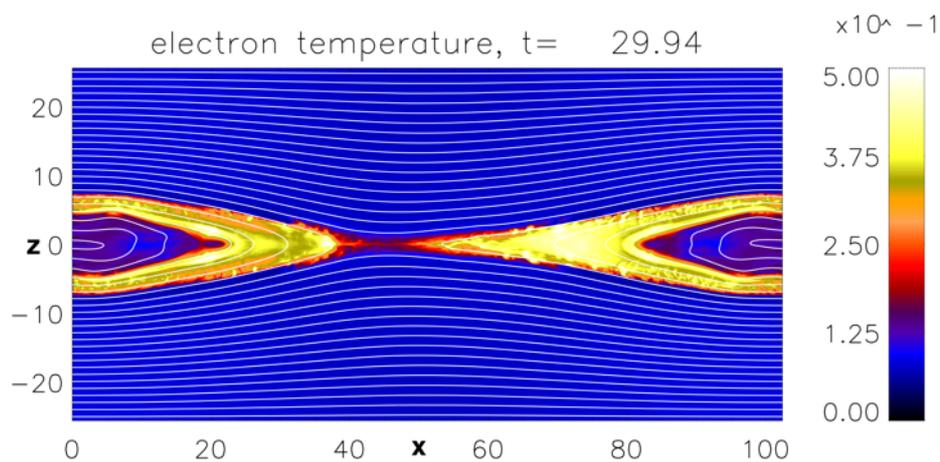

Figure 9: Global plot of the electron temperature. The figure shows a strong temperature increase at or just inside the separatrix. The colder electrons inside the island across the periodic boundaries are the remnants of the initial condition, i.e., electrons pushed to the side by the reconnection process.

Figure 9 supports the view that some process, located at the separatrix in the vicinity of the X-line, and perhaps slightly further inside the outflow region further away, should be playing a role. We see in this overview picture that there is a clear difference in electron temperature between the inflow and outflow regions. The transition between the two temperatures appears to be co-located with regions of turbulence, another indication that turbulent processes may play an important role here.

This view is further supported by inspection of Figure 10, which displays the electron temperature zoomed into the rectangle shown in fig. 1. It is very evident that there is a very close spatial correlation between the electrostatic structures shown in fig. 5 and strong electron temperature enhancements. Therefore, it appears warranted to quantify this effect in some suitable manner.

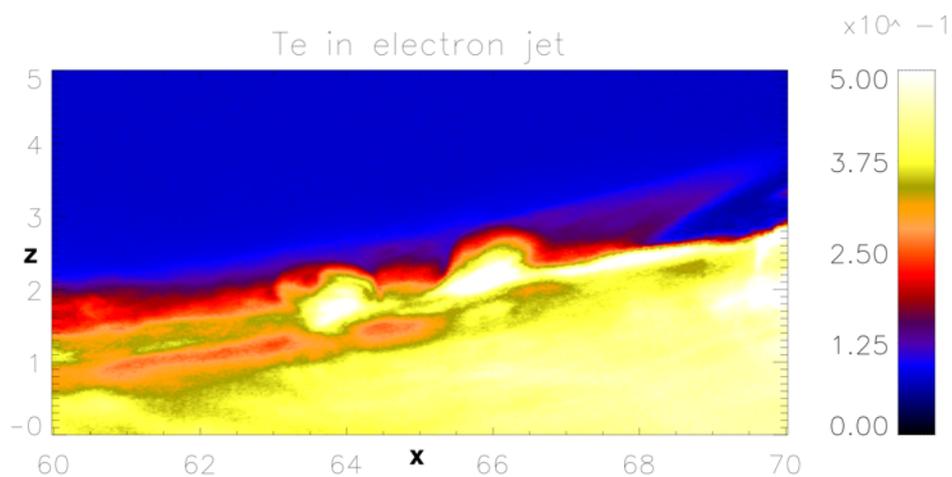

Figure 10: Zoomed-in view of the electron temperature. There are very noticeable temperature enhancements in association with the velocity shear layer, and, in particular, with the double structures of the electric field.

Similar to a recent investigation[7], we quantify this process by means of integrating the electron energy equation:

$$\frac{\partial p}{\partial t} = -\nabla \cdot (\vec{v}p) - \frac{2}{3}\sum_{l} P_{ll}\frac{\partial}{\partial x_l}v_l - \frac{1}{3}\sum_{l,i}\frac{\partial}{\partial x_i}Q_{lii} - \frac{2}{3}\sum_{\substack{l,i \\ l\neq i}} P_{li}\frac{\partial}{\partial x_i}v_l \qquad (2)$$

Here $p$ denotes the trace of the electron pressure tensor, $P_{ij}$ the components of the electron pressure tensor, $Q_{ijk}$ the heat flux triple-tensor, $x_l$ the three coordinate directions, and $v_l$ the corresponding velocity components. The pressure changes due to compression, expansion, and convection are represented by the first two terms on the right-hand-side of (2) – with the second term correcting the first for anisotropy. Heat flux effects, which are essentially a correction to the first two terms for complex distributions[7], are captured by the third term. The last term in (2) involves cross-derivatives of the flow velocity, which resemble viscous contributions in a collisional system, and off-diagonal pressure. These terms are referred to as "quasi-viscous." Similar to our previous analysis of the electron diffusion region[7], we achieved results of sufficient fidelity by reducing the time step from $\omega_{pe}dt=0.25$ to $\omega_{pe}dt=0.01$ during the period of analysis. The heat flux was averaged over outputs produced at full plasma period intervals between $\Omega_i t=29.9$ and $\Omega_i t=29.95$, lower order moments were averaged over outputs produced at full plasma period intervals between $\Omega_i t=29.915$ and $\Omega_i t=29.935$.

Analyzing the terms in (2) locally does not lead to conclusive information due to the relatively high fluctuation level. In order to obtain a global balance, it is best to integrate over some appropriately chosen volume to average out fluctuations and to obtain a net effect. Since we expect significant contributions in the region around the separatrix, an appropriately chosen integration region should be bounded by flux tubes. We hence approach the problem from this angle: we determine the flux function A, for which:

$$\vec{B} = -\nabla A \times \vec{e}_y + B_y \vec{e}_y \tag{3}$$

We normalize $A$ such that $A=0$ at outermost flux tubes at the $z=z_{max}$ boundaries. The minimum value of A denotes the center of the island structure at the periodic boundary, located at approximately $x=0$ (or $x=x_{max}$). If $A_0$ denotes a field line in the inflow region, the condition

$$A > A_0 \tag{4}$$

defines: an area in the *x-z* plane, which is, on one side, bounded by the upper ($z=z_{max}$) boundary, and on the other by the field line, for which $A=A_0$, in the upper inflow region, and a corresponding area between the lower boundary and the field line, for which $A=A_0$, in the lower inflow region. As soon as $A_0$ denotes outflow flux tubes, the two areas connect and are now bounded by the two outflow field lines, for which $A=A_0$. We can now integrate (2) over this volume, and then vary $A_0$ to investigate whether there are significant changes around a specific field line. The result of this integration is shown in Figure 11.

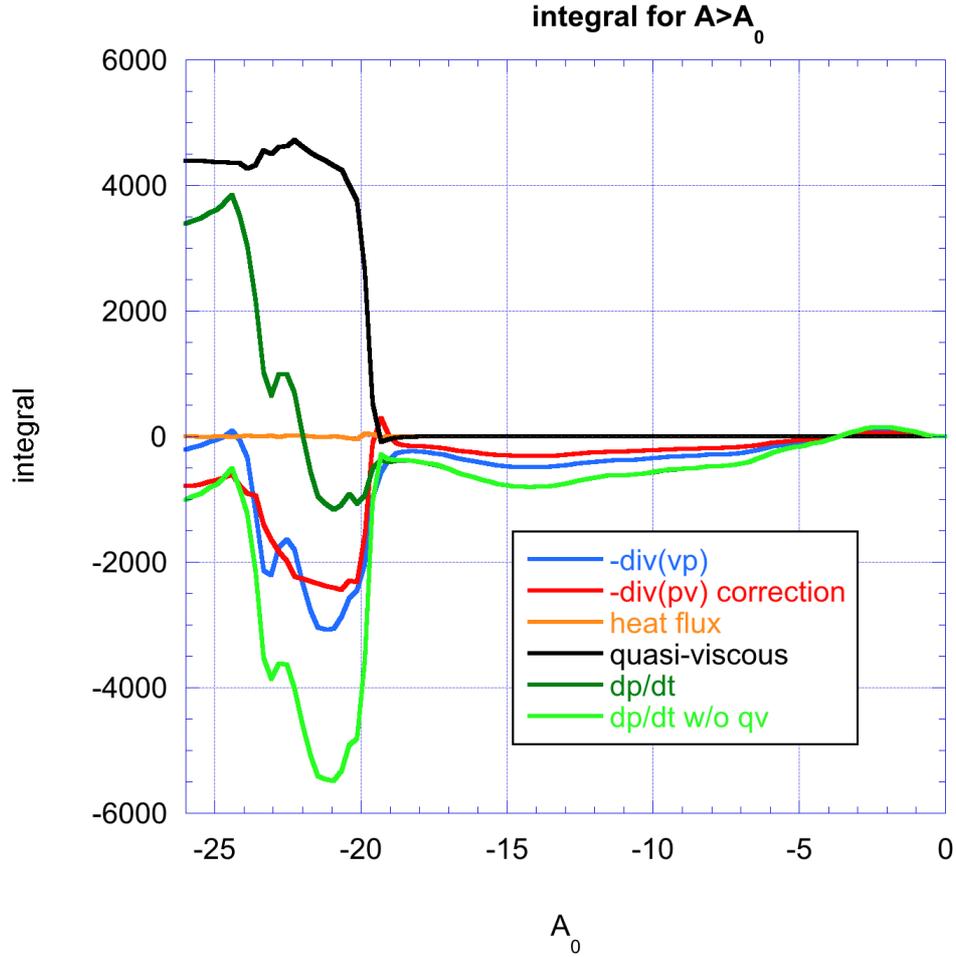

Figure 11: Integration of the various terms of the energy equation over a volume bounded by flux tubes at $\Omega_i t=29.94$. The figure shows that the quasi-viscous contribution is the main energy source. It becomes important as soon as the integration volume extends past the separatrix field line.

The integration of the heat flux shows (orange) the expected very small contribution: the periodic system investigated here combined with the field line boundary renders global heat flux effects negligible. We note that, in principle, the contribution could be larger and negative in an open system. The compression-convection terms (blue and red), which are the first two terms on the right-hand-side of (2), show very significant contributions. Above the separatrix field lines, located approximately at $A_0 \approx -19.5$, these terms indicate a slight reduction of the thermal energy content of flux tubes in the inflow region, which is consistent with the reconnection-generated expansion in this region. If we move the boundary into the reconnected flux region, we find that these terms alone would first lead to a strong reduction of the thermal energy density. While this negative contribution might seem surprising, it is merely the consequence of the convection of colder and low-density electrons from the lobe across the separatrix. In the absence of additional heating, we would expect a local reduction of thermal energy density due to this effect.

A bit further inside the reconnected flux region, at $A_0 \approx -21$, compression effects, generated by trailing reconnection flows and magnetic flux, begin to contribute to increasing thermal energy densities. This increase continues until $A_0 \approx -24$, at which point we have reached those flux tubes, which threaded the current layer in the initial configuration. From here on to the center of the island across the periodic boundary at $A_0 \approx -26$, further compression-convection effects are small. Their combination, indicated by the light green line in fig. 11, would lead to an overall reduction in thermal energy, an effect, which would be in contradition to the expected conversion of magnetic energy to particle energy. We note that this reduction effect could be even larger in an open system, where compression effects by back-pressure would naturally be even smaller.

The term restoring the proper energy conversion is the quasi-viscous contribution. Its effects are represented by the black line in fig. 11. It features very small effects in the inflow field line regions and deeper inside the outflow regions. In the transition around the separatrix, however, this term has dramatic effects: it essentially accounts for the entire electron energization between the inflow and outflow, leading to a positive overall thermal energy change (green curve). Similar to results pertaining to the electron diffusion region[7], we thus find that quasi-viscous effects are again a critical contributor to the energy balance, even on much larger scales.

Intuitively, we expect that this quasi-viscous contribution should be connected to the instabilities discussed above due to their co-location. In order to investigate whether such a relation exists and what its nature might be, we integrate the individual terms of the quasi-viscous heating term

$$H_{qv} = -\frac{2}{3}\sum_{\substack{l,i \\ l \neq i}} P_{li}\frac{\partial}{\partial x_i}v_l = -\frac{2}{3}\left(P_{xz}\frac{\partial}{\partial z}v_x + P_{xy}\frac{\partial}{\partial x}v_y + P_{yz}\frac{\partial}{\partial z}v_y + P_{xz}\frac{\partial}{\partial x}v_z\right) \quad (5)$$

in the same way as before. The result is displayed in Figure 12.

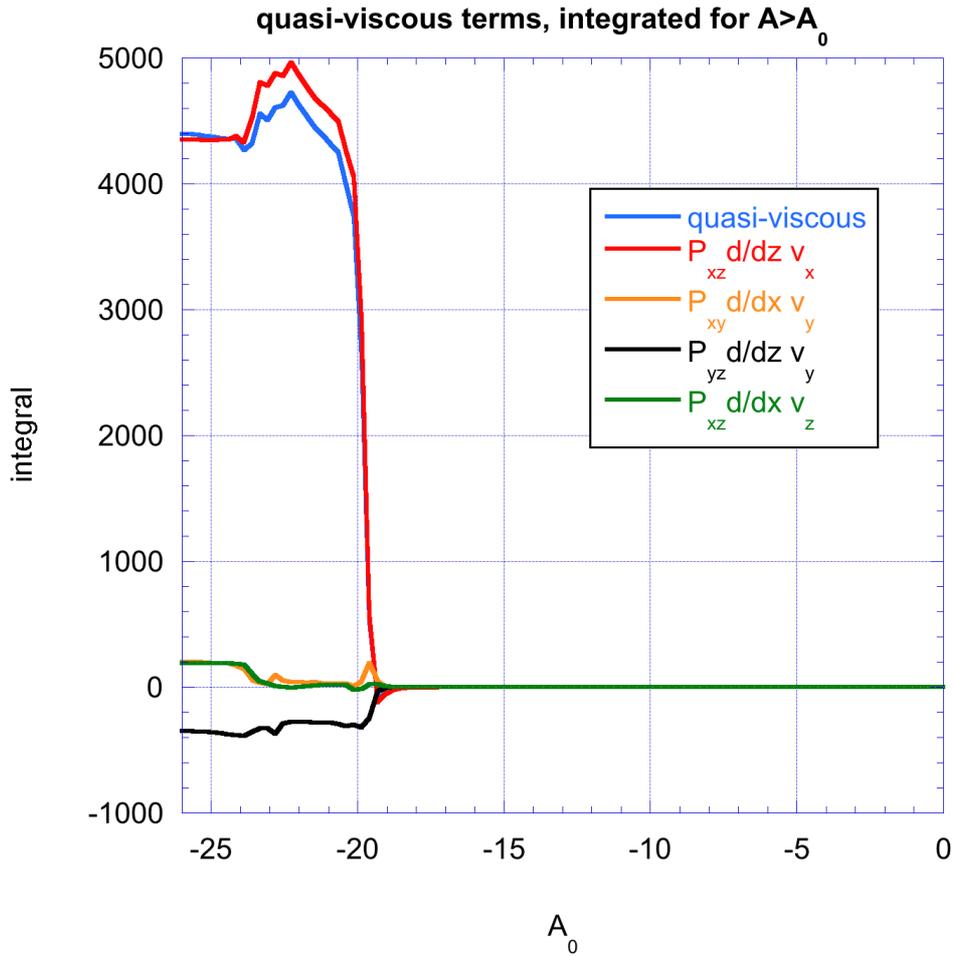

Figure 12: Integration of the components of the quasi-viscous heating term $\Omega_i t=29.94$. The dominance of the term $\sim P_{xz}\frac{\partial}{\partial z}v_x$ shows that the heating is indeed related to velocity shear effects. The negative contributions of the term $\sim P_{yz}\frac{\partial}{\partial z}v_y$ at the separatrix dominate, by far, over the small positive contribution this term has in the electron diffusion region[7]. In the absence of instabilties the contribution $\sim P_{xz}\frac{\partial}{\partial z}v_x$ would likely be reduced, and be balanced by that $\sim P_{yz}\frac{\partial}{\partial z}v_y$.

Figure 12 shows that, of these terms, the one proportional to $P_{xz}\frac{\partial}{\partial z}v_x$ dominates by far, indicating that effects related to the velocity shear appear to account for the heating. However, it cannot be just the velocity shear in itself, which leads to heating: instabilities are critical. In the absence of instability, we would just see statically overlapping beams on gyro-scales, and even a static $P_{xz}$ component, but with no further interaction. Even though overlapping beams with Larmor-scale gradients between them can locally, in a region of Larmor-radius width, generate an apparently larger temperature, these structures are static and therefore there is no net heating effect as no beam energy is being dissipated. Instead, we expect that the positive contribution associated with $P_{xz}\frac{\partial}{\partial z}v_x$ will be balanced by a negative contribution associated with $P_{yz}\frac{\partial}{\partial z}v_y$ so that no net heating results. We note that the latter term is also negative even in the presence of instabilities, as shown in fig. 12.

In summary, we find that quasi-viscous heating effects are the main contributor to electron energization from the inflow to the outflow in the present model. As expected intuitively, beam-beam instabilities rendered unstable by very strong velocity shears are the main contributor to this heating effect.

## V. SUMMARY AND DISCUSSION

In this paper, we described an investigation into the role instabilities at the reconnection outflow separatrix can play in heating the electron population in the outflow region. Electron heating in the transition from reconnection inflow to outflow is one important piece of the energy conversion puzzle in collisionless magnetic reconnection: which processes replace the shocks or discontinuities, which facilitate energy conversion in MHD plasmas?

Instabilities in this general region have been seen before in a number of different investigations[16-21], and their possible significance in this context warranted investigation. In an in-depth analysis of our simulation model, we found a beam-beam instability to be operating at the interface between oppositely directed electron jets in the separatrix region. This instability was found to be enabled by finite Larmor radius-effects, which can generate distributions unstable to a primarily electrostatic instability if the gradient scale length between the opposing beams is small enough. We demonstrated the existence of unstable, multi-component, electron distributions in the region of strong, bi-polar, electric fields with substantial components parallel to the magnetic field. The electrostatic solitary structures contained trapped electron distributions, which moved with an average velocity of approximately 50% of the electron Alfvén speed. Contrary to earlier investigations[19] we did not see significant whistler wings, the absence of which we attributed to the relatively slow propagation speed of the solitary structures.

These moving structures generated significant perturbations in the average electron flow. Vortical flows associated with them served to further enhance the mixing of electrons from opposing beams. A temperature analysis revealed a close co-location of this type of turbulence and a substantial temperature increase moving from the inflow regions to the outflow jet – an impression that was supported by inspection of the local temperature variations. In order to analyze the way this temperature increase was created, we performed an integration over different terms of the electron energy equation. The integration region was chosen to be bounded by magnetic field lines, and the bounding field line was varied to investigate the existence of a critical heating region.

We found no significant heat flux contributions and attributed this absence to the periodicity of the simulation model. Compression and convection terms, however, appeared to contribute negatively to the overall electron energy density: a strong negative contribution near the separatrix was found to be only partially compensated for by a positive contribution in the region, where reconnected flux tubes can get compressed between the magnetic island on one side, and the reconnection outflow on the other. The convection-compression terms alone would thus lead to a negative net effect when integrated over the entire simulation volume.

The only net contribution to electron heating was found to be the quasi-viscous term. It showed a strong contribution in the area around the separatrix, which suggested its association with the turbulent features discussed earlier. This association was confirmed in a detailed analysis of the individual terms of the quasi-viscous contribution. Here we found that the term, which involves the derivative of the electron bulk flow normal to the bulk flow direction, provided, by far, the largest contribution. We argued that a purely static velocity gradient cannot, even when sharp, account for any heating effects. Therefore, we concluded that instabilities of the type seen in our model can provide critical electron heating.

An interesting question is related to the energy source of the electron heating. Clearly, the immediate source is the counter-streaming flows. However, the instability responsible for the heating tends to broaden the velocity gradient and, hence, change the flow profile. The electron flows, however, are parts of the current system, which is associated with the quadrupolar guide field. Any process reducing the structure of these currents will thus generate electric fields, which counteract the effect of the instabilities. It seems conceivable that these electric fields alone, i.e., in the absence of other effects, would also reduce the guide field. This process chain, which is conceptually similar to processes inside the electron diffusion region[7], could thus indirectly convert magnetic energy stored in the guide field to electron thermal energy. A detailed investigation of this mechanism is outside the scope of the present analysis but will be pursued in the future.

Finally, we note that there are likely to be other instabilities, which could have similar effects, and that our simulation is, like most kinetic models, still fairly limited in physical dimensions. In particular, it is conceivable that additional instabilities can arise from the interaction of multiple beams. Nevertheless, the present results indicate one possible solution to the inflow-to-outflow heating problem.

**ACKNOWLEDGEMENTS**

This work was funded by NASA's MMS project, and in part by grants from NASA 80NSSC18K0754, DOE DESC0016278 and NSF AGS1619584, AGS-1202537, AGS-1543598, and AGS-1552142. Work at the University of Bergen was supported

by the Research Council of Norway/CoE under contract 223252/F50. The authors recognize the tremendous effort in developing and operating the MMS spacecraft and instruments and sincerely thank all involved.